# Scintillator-Based Electromagnetic Calorimeter Prototype and Beam Test Results at FNAL

Adil Khan For the CALICE collaboration

1 – Center For High Energy Physics Department
Kyungpook National Univeraity, Daegu – S.Korea

A prototype Scintillator-Tungsten electromagnetic calorimeter (ScECAL) for the ILC detector was tested in 2008 at the Fermilab test beam. Data were collected with electron, pion and muon beams in the energy range 1 to 32GeV combined with hadronic calorimeter and Tail catcher. One of the main objectives of the CALICE program is to establish the technology of Scintillator-based electromagnetic calorimeter and validate the prformance of the calorimeter. From preliminary results of the first approach of analysis with electron beam, we obtain the ScECAL energy resolution $\sigma_{stochastic} = 15.15 \pm 0.03\%$ and $\sigma_{constant} = 1.44 \pm 0.02\%$. The deviation from the linear response is calculated to be less than 6%.

## 1 Introduction

The International Linear Collider (ILC) is a future $e^+e^-$ collider experiment that operates with a center-of-mass energy 220-1000 GeV. The main goals of the ILC are the precise measurement of the Higgs boson and SUSY particles as well as precise tests of the Standard Model using W, Z bosons and the top quark. Since these important physics processes will decay predominantly to multi-jet final states, the utmost target of the detector performance is the jet energy resolution. The goal of the jet energy resolution is set to $\frac{\sigma_E}{E} \sim 30\%/\sqrt{E}$ [1]. The Particle Flow Algorithm (PFA) is adopted to achieve this jet energy resolution[2]. In The PFA approach every particle in the final state is reconstructed. Each individual particle in a jet is separated into charged or neutral particle, and measure the jet energy by combining momentum of charged particles measured by tracker and energy of neutral particles by the calorimeter. In order to realize the PFA, the main aim of the calorimeter at the ILC is to separate and identify each particle in a jet with fine segmentation.

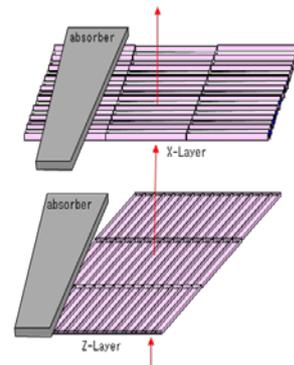

Figure1.Schematic view of the Scintillator-strip calorimeter.

To achieve such fine granularity with a sandwich-type sampling calorimeter, the design of ScECAL consists of 3 mm thick tungsten layers as absorber and Scintillator strip layers as shown in Figure 1. Size of individual Scintillator strip is 1 x 4.5 cm with thickness of 2 mm. The Scintillator strips are orthogonal in successive layers to achieve effective 1x1 cm cell size. Signal from each individual strip is read by a novel type semiconductor photo-sensor, called Multi Pixel Photon Counter (MPPC)[3]. The MPPC is attached at the end of each Scintillator strip. In this paper we report on development of the Scintillator strip electromagnetic calorimeter (ScECAL) prototype which is one of the ongoing activities in the CALICE collaboration[4].

## 2 The Scintillator Strip Calorimeter Prototype

An electromagnetic calorimeter (ECAL) test module with an extruded Scintillator-strip structure has been constructed and tested. Figure 2 shows the structure and picture of the ECAL prototype fabricated.

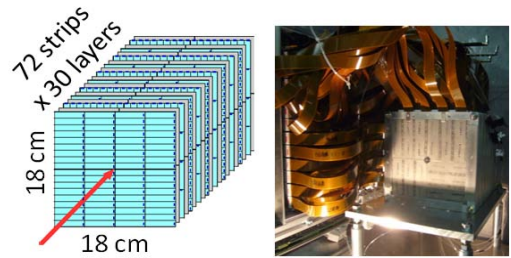

Figure 2. (Left) Schematic structure of the ScECAL. Red arrow indicates beam direction and (right) the ScECAL prototype.

The module consists of 30 pairs of Scintillator and Tungsten as absorber layers of thickness 3 and 3.5 mm respectively. The size of each extruded Scintillator strip is 10 mm x 45 mm x 3 mm. As shown in Figure 3, each Scintillator strip is totally wrapped by a reflector film produced by Kimoto Co. to optically isolate them from each other and to increase amount of light conducted by the MPPC.

The scintillation light produced in plastic Scintillator strips enter the wavelength shifting(WLS) fiber placed into the center hole of plastic Scintillator which are guided to the sensitive photo detector 1600 pixel MPPC (Multi Pixel Photon Counter) with a sensitive region of 1 x 1 $mm^2$. In order to get the best performance, great care should be taken in exact matching of the position of WLS fiber to the surface of photon sensor. The flat cable is used as a bridge for directing the signal from MPPC to the readout electronic baseboard.

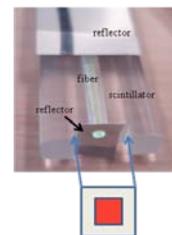

Figure 3.Picture of the Scintillator strip

Each active layer is a matrix of 72 Scintillator strips. In successive alternate layers Scintillator strips are orthogonal to achieve 1 cm granularity both in two perpendicular (X and Y) directions. Overall size of the ScECAL is 18 cm $\times$ 18 cm $\times$ 20 cm with a total length equivalent to 21 radiation lengths (21X0).

## 3 Beam test at Fermilab and Preliminary Results

The ScECAL beam tests were performed in September 2008 and May 2009 at the Fermilab meson test beamline (MT6 section B). These tests were carried out with a combined system of ScECAL, followed by a hadronic calorimeter (AHCAL)[5] and then a Tail catcher (TCMT)[6], the latter two both built from a Scintillator-iron sandwich structure as shown in Figure 4(right). One of the main goal of the beam test is to establish the Scintillator-based ECAL technology by evaluating the performance of the ScECAL prototype.

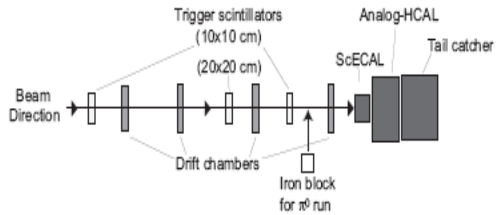 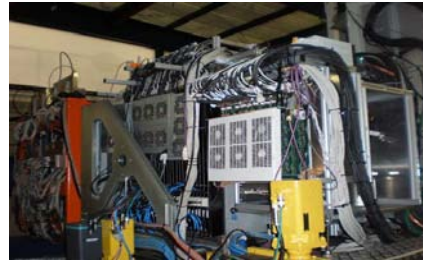

Figure 4.(left) Schematic view of beamline setup. (right) Detector side view showing ScECAL+AHCAL+TCMT

The layout of the CALICE calorimeters in the test beamline is shown schematically in figure 4(left). Upstream of the calorimeters there are four layers of drift chambers for beam position measurement. Signal from 10×10 cm or 20×20 cm trigger Scintillators are used for data taking trigger. We have used electron, charged pion and muon beams with energy range of 1-32 GeV.

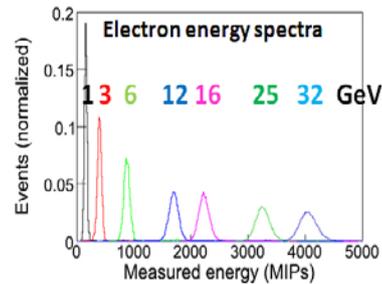

The first look into data has been started by the calibration of the strip response using Minimum Ionizing Particle (MIP) signal by muon beam. Therefore the measured energy is expressed in unit of MIP equivalent. After the calibration, energy measurement of the electron beam with various energies has been done. The energy spectra of electrons measured by the ScECAL are displayed in Figure 5.

Figure 5. Energy spectra of electrons measured on the ScECAL

Preliminary results of the energy resolution and linearity are obtained for the center and uniform region of the detector are shown in Figures 6(left) & 6(right) respectively. There is no significant difference between the energy resolution and linearity for the center and uniform region. Detailed analyses to evaluate energy resolution and linearity response are currently underway.

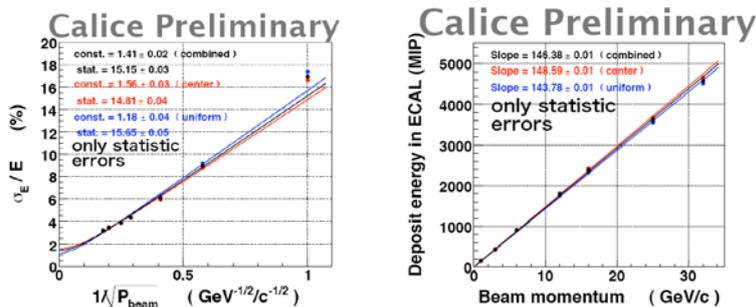

Figure 6.(left) The resolution of ScECAL response of energy as a function of beam momentum. (right) linearity response of measured deposited energy in ScECAL as a function of beam momentum .

The data points are fitted with a formula $\frac{\sigma_E}{E} = \sqrt{\frac{\sigma_{stat}^2}{E} + \sigma_{const}^2}$, where $\sigma_{stat}$ and $\sigma_{const}$ denote stochastic and constant terms of the energy resolution, respectively. As a result of the fit, we obtain σ$_{stochastic}$ = 15.15±0.03% and σ$_{constant}$ = 1.44±0.02% (errors are statistical only) which are reasonably good for a preliminary result.

## 4 Conclusion

We have built the ScECAL second prototype and performed a beam test combined with the AHCAL + TCMT system using $e^-$, $\pi^-$ and $\mu^-$ beams. In energy range of 1-32 GeV & have observed a clean electron energy spectra measured with the ScECAL. The test also gives us a lot of experience with the whole system including operation of the detector, electronics and various monitoring system. Although analyses of the collected data are still extensively ongoing, preliminary results are promising and demonstrate the feasibility of the ScECAL.

## 5 Acknowledgements

The author would like to thank to Misung Chemical Co., Daegu, Korea, for producing Scintillator strips with a good quality. This work has been carried out with the support of the World Class University project(WCU) by Korea National Research Foundation. Many people contribute to this project. In particular, I'd like to express special thanks to Fermilab accelerator staffs to provide a good quality of beam. This work is also supported in part by the Creative Scientific Research Grants No.~18GS0202 of the Japan Society for Promotion of Science (JSPS).

## References

[1] Letter of Intent (ILD). Available:http://www.ilcild .org/documents/ild-letter-of-intent/LOI.pdf/view.
[2] M.A. Thomson, arXiv.physics/060726 (2007).
[3] S. Uozumi, Nuclear Instrument and methods A 581 427 (2007).
[4] https://twiki.cern.ch/twiki/bin/view/CALICE/CaliceCollaboration.
[5] N. Meyer, et al, (CALICE collaboratio), CAN-014, 5pp, (2008).
[6] A. Dyshkart, et. Al, Tail Catcher moun Tracker for the CALICE test beam, AIP conf. proc. 867 pp592-599, (2006).